\documentclass[a4paper]{PoS}

\usepackage[numbers]{natbib}

\title{Evaluating nuclear physics inputs in core-collapse supernova models}

\ShortTitle{Evaluating nuclear physics inputs in core-collapse supernova models}

\author{\speaker{Eric J. Lentz}$^{ab}$, W. Raphael Hix$^{ba}$, Mark~L. Baird$^{ca}$, O.~E.~Bronson Messer$^{da}$, and Anthony Mezzacappa$^{ba}$\thanks{
This work was supported by the NSF PetaApps program under grant  OCI-0749242; by NASA under grant 07-ATFP07-0011; and by the Office of Nuclear Physics and Office of Advaced Scientific Computing Research, U. S. 
Department of Energy.
This research used resources of the Oak Ridge Leadership Computing Facility at the Oak Ridge National Laboratory provided through the INCITE program.
This research was supported by an allocation of advanced computing resources provided by the National Science Foundation.
The computations were performed on Kraken (a Cray XT5) at the National Institute for Computational Sciences.
}\\
\llap{$^a$}Department of Physics \& Astronomy, University of Tennessee\\ 
   Knoxville, TN 37996, USA\\
\llap{$^b$}Physics Division, Oak Ridge National Laboratory\\
   Oak Ridge, TN 37831, USA\\
\llap{$^c$}Nuclear Science \& Technology Division, Oak Ridge National Laboratory\\ 
   Oak Ridge, TN 37831, USA\\
\llap{$^d$}Oak Ridge Leadership Computing Facility, Oak Ridge National Laboratory\\ 
   Oak Ridge, TN 37831, USA\\

           E-mail: \email{elentz@utk.edu}}

\abstract{Core-collapse supernova models depend on the details of the nuclear and weak interaction physics inputs just as they depend on the details of the macroscopic physics (transport, hydrodynamics, etc.), numerical methods, and progenitors. 
We present preliminary results from our ongoing comparison studies of nuclear and weak interaction physics inputs to core collapse supernova models using the spherically-symmetric, general relativistic, neutrino radiation hydrodynamics code Agile-Boltztran.
We focus on comparisons of the effects of the nuclear EoS and the effects of improving the opacities, particularly neutrino--nucleon interactions. 
}

\FullConference{11th Symposium on Nuclei in the Cosmos, NIC XI\\
		July 19-23, 2010\\
		Heidelberg, Germany}

\begin{document}

\section{Stellar core collapse and model inputs}

The evolution of massive stars ($M \gtrsim 10\ M_\odot$) concludes when Si-burning builds a core of Fe-peak elements that can no longer be supported by electron degeneracy pressure and begins to collapse.
During collapse, electron capture (EC) on nuclei and free nucleons reduces the electron fraction, $Y_e$, in the Fe-core until the released neutrinos become trapped in the increasingly neutrino-opaque core.
When the collapsing core reaches super-nuclear densities, $\rho_c \sim 3 \times 10^{14}$~g~cm$^{-3}$, nuclear repulsion halts further compression, and the core rebounds to form a shock, $M_{\rm sh} \propto Y_e^2$.
The outgoing shock dissociates the heavy nuclei it passes, which, along with losses to neutrino emission, saps the energy of the rebound shock, and it stalls above the newly-formed proto-neutron star (PNS).
This standing accretion shock (SAS) is likely revived by heating from $\nu_e$ and ${\bar \nu}_e$ absorption on free nucleons behind the shock, to disrupt the star as a supernova.
Neutrino heating depends on the luminosity and thus on the opacities in the neutrino emitting region below.
Multidimensional effects, including neutrino-driven convection and the SAS instability, expand the heating region and enhance the heating that drives the explosion.
This paper is a preview of studies on the effects of changes in neutrino opacities \cite{opac} and the nuclear equation of state (EoS) \cite{eos} on the development of Fe-core collapse supernovae using Agile--Boltztran \cite{MeBr93a,LiMeMe04,opac} --- a spherically symmetric, general relativistic, neutrino radiation hydrodynamics code.
All of our models have 102 adaptive mass shells; 20-energy-group, 8-angle-quadrature transport with updated opacities for all $\nu$ species; LS~180 EoS \cite{LaSw91}; and an up-to-date $15\ M_\odot$ progenitor  \cite{WoHe07} unless otherwise stated.

\section{Neutrino opacities}

The  often used opacity set defined by \citet{Brue85}  includes emission, absorption, and scattering on free nucleons and nuclei, scattering on electrons, and the pair source $e^+e^-  \leftrightarrow \nu\bar{\nu} $.
Additional interactions such as nucleon--nucleon bremsstrahlung, $ NN \leftrightarrow NN\nu\bar{\nu}$ (where $N$ is any free nucleon) \cite{HaRa98}, and neutrino-pair flavor conversion, $\nu_\alpha \bar{\nu}_\alpha \leftrightarrow \nu_\beta \bar{\nu}_\beta$ \cite{BuJaKe03}, have been shown to be important sources of $\mu\tau$-neutrinos.
Replacing the independent particle model (IPM) for EC on nuclei  with an  emission table using detailed nuclear data \cite{LaMa00,LaMaSa03} (LMSH EC table) changes de-leptonization during collapse and the $\rho$ and $Y_e$ gradients that the shock traverses \cite{HiMeMe03}.
Improved interactions of neutrinos with free nucleons  that include nucleon recoil, dense matter correlations, etc.\ \cite{BuSa98,RePrLa98} and corrections for weak magnetism \cite{Horo02} reduce the $\nu$--nucleon opacities and provide for spectral change not available in the \citet{Brue85} $\nu$--nucleon opacities.
For these studies we have added to the opacity set from \cite{Brue85} the NN Bremsstrahlung, LMSH EC table, and enhanced $\nu$--nucleon opacities.

\begin{figure}
\begin{center}
\includegraphics[width=.65\textwidth]{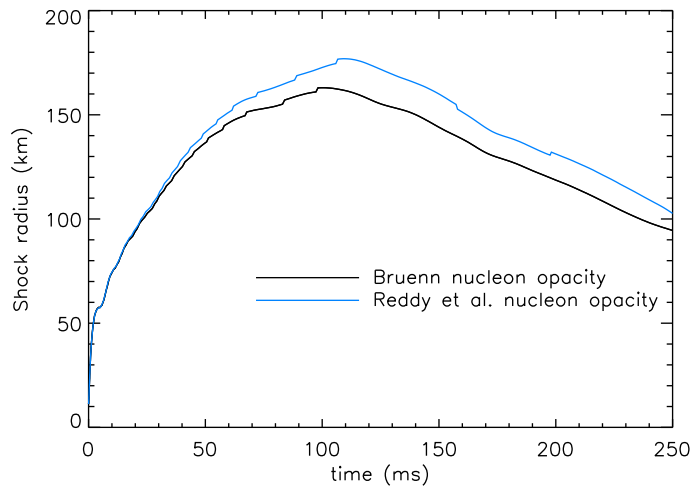}
\end{center}
\caption{Shock radii for $15\ M_\odot$ models \cite{WoHe07} with different nucleon opacities plotted against time after core bounce.
The blue line is the updated opacity set that includes the \citet{RePrLa98} $\nu$--nucleon interactions.
The black line uses the \citet{Brue85} $\nu$--nucleon opacities.
All other inputs and parameters are the same.
The small jumps in shock radius result from the shock tracking algorithm shifting to an adjacent zone.}
\label{fig:opac}
\end{figure}

We have implemented the \citet{RePrLa98} $\nu$--nucleon opacities using an energy sub-grid without weak magnetism corrections and computed  models with all permutations of the Reddy and Bruenn $\nu$--nucleon opacities.
We find that the shock radius (Fig.~\ref{fig:opac}) is larger for models using the Reddy $\nu$-opacities (blue) than for the Bruenn $\nu$-opacities (black).
The differential effect is larger for the change in scattering opacity than for the change in emission/absorption opacity by a factor of 3--4 depending on the epoch \cite{opac}.
The luminosities of $\nu_e$ and $\bar{\nu}_e$ both increase with the smaller \citet{RePrLa98} opacities, which allow neutrinos to stream more easily from the PNS while their r.m.s. energies remain largely unchanged as measured outside the shock.
The difference between the increased luminosity from scattering and emission/absorption changes, respectively, parallels the changes in shock radius, with the larger effect resulting from updated scattering.
Both the increased luminosity and shock radii with updated nucleon interactions \cite{BuSa98} were seen in the 1-D models of \citet{RaBuJa02}.

\section{Nuclear equation of state}

\begin{figure}
\begin{center}
\includegraphics[width=.6\textwidth]{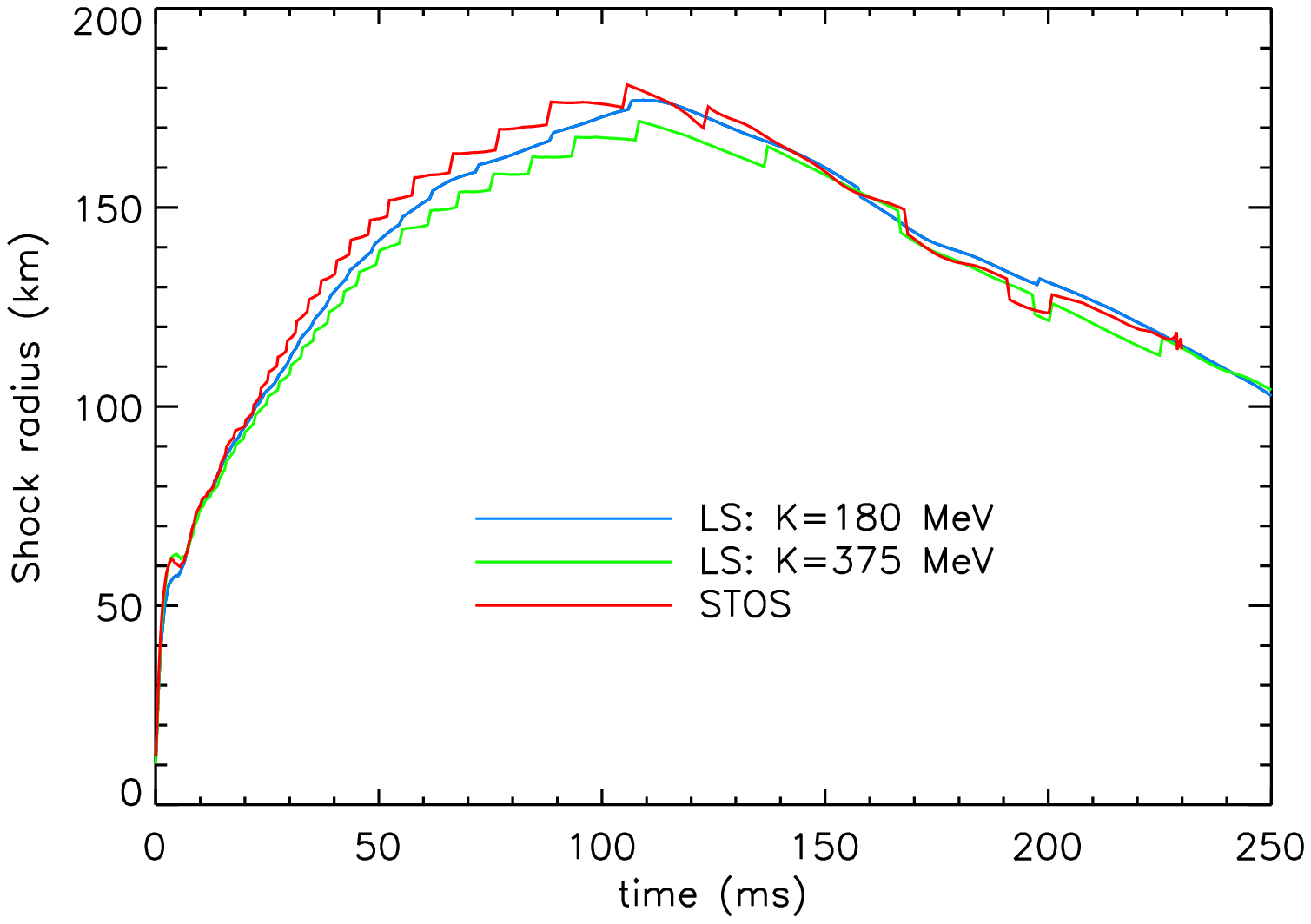}
\includegraphics[width=.98\textwidth]{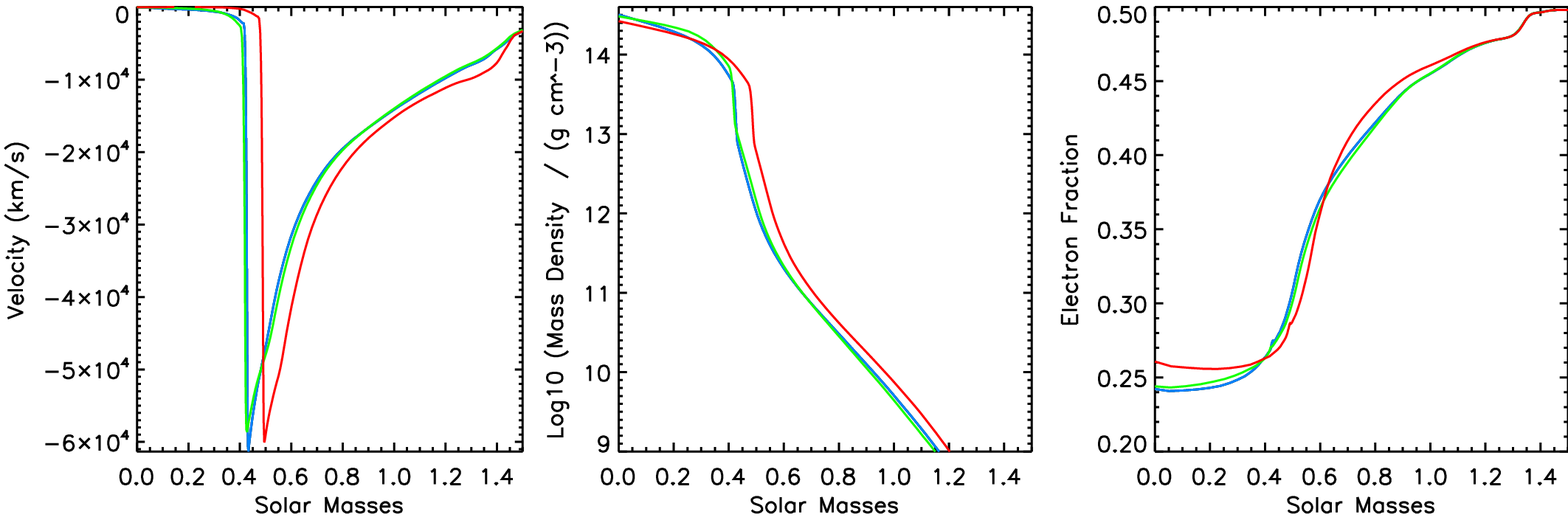}
\end{center}
\caption{ {\it Upper panel}: Shock radii for $15\ M_\odot$ models \cite{WoHe07} with different nuclear EoSs plotted against time after core bounce.
{\it Lower panel}: Velocity (left), density (center), and $Y_e$ (right) profiles for the models at core bounce.
The models use the \citet{LaSw91} EoS with $\kappa_s = 180$~MeV (blue, same as the blue line in Fig.~1) and  $\kappa_s = 375$~MeV (green),  and the \citet{ShToOy98} EoS (red).
}
\label{fig:eos}
\end{figure}

The nuclear EoS determines the thermodynamic properties of the fluid in SN simulations in two dynamically important regions---the high-density nuclear matter in the PNS and the nuclear statistical equilibrium (NSE) region at intermediate densities.
Outside the Fe-core we include an optically thin non-NSE region with a Si/Fe mixture.
We have computed simulations using two EoSs---the liquid-drop model of \citet{LaSw91} (LS) EoS with a symmetry energy, $E_{\rm sym} = 29.3$ MeV and bulk incompressibility, $\kappa_s=180$ and 375~MeV, and the relativistic mean field model of \citet{ShToOy98} (STOS) EoS with $E_{\rm sym} = 36.9$ MeV and $\kappa_s = 281$~MeV.
A comparison of models using the LS~EoS shows little variance with $\kappa_s$ in initial shock position ($\sim 0.01\ M_\odot$, Fig.~\ref{fig:eos}, lower left) or core density ($\sim 5\%$, lower center) at bounce.
At later times, the softer, more compressible, LS~180 EoS model with a more compact core had luminosities up to 10\%  larger (similar to the change from updated scattering) than the model with the stiffer LS~375 EoS; and a small, detectable, change in the maximal extent of the shock (Fig.~\ref{fig:eos} upper).
A similarly small difference due to $\kappa_s$ in LS EoS models was seen by \citet{SwLaMy94}.

The differences at bounce between the models using the LS and STOS EoSs  originate primarily in the NSE region during collapse where the STOS EoS gives a lower abundance of heavy nuclei, resulting in less EC by nuclei \cite{HiLeBa08} and therefore a larger initial shock position ($0.49\ M_\odot$ versus $0.44\ M_\odot$ for the LS~180 EoS model, Fig.~\ref{fig:eos} lower).
The differences in shock progression (Fig.~\ref{fig:eos}, upper) are comparable to, but smaller than, those seen by \cite{JaBuKi05,HiLeBa08,Mare07}, who used the LMSH EC as we did, but larger than the differences seen by \cite{SuYaSu05} who did not and therefore $Y_e$ reduction during collapse was dominated by capture on free nucleons, the abundances of which are affected by  $E_{\rm sym}$.
Each of the referred simulations used a different $15\ M_\odot$ progenitor and a different opacity set which complicates the comparison.
We also found that the differences  for variations in EoS using the $12\ M_\odot$ progenitor \cite{WoHe07} are even smaller than they are for the $15\ M_\odot$ progenitor \cite{eos}. 

\section{Conclusions}

Improved $\nu$--nucleon interactions \cite{BuSa98,RePrLa98} and EC on nuclei \cite{LaMaSa03} show definite, and complex, effects on the evolution of the collapsed core and should be included in all realistic supernova simulations.
For example, we have shown here nontrivial and {\it consistent} enhancements in the shock radius by using the \citet{RePrLa98} $\nu$--nucleon scattering opacities.
Variations in the compressibility of nuclear matter within the LS EoS seem to have little effect on shock progression through the times  when multidimensional effects would start.
Switching between the LS and STOS EoSs leads to larger differences than switching compressibilities within the LS EoS set, but these differences remain smaller than those introduced by changing the neutrino opacities.
Finally, the differences depend also on the progenitor used.
Thus a systematic and comprehensive study must traverse the three axes of opacity, EoS, and stellar progenitor.
Such a study is forthcoming \cite{opac,eos}.

\bibliographystyle{apsrev}
%\begin{thebibliography}{99}
\bibliography{apj_journals,supernova,supersupl}

\begin{thebibliography}{22}
\expandafter\ifx\csname natexlab\endcsname\relax\def\natexlab#1{#1}\fi
\expandafter\ifx\csname bibnamefont\endcsname\relax
  \def\bibnamefont#1{#1}\fi
\expandafter\ifx\csname bibfnamefont\endcsname\relax
  \def\bibfnamefont#1{#1}\fi
\expandafter\ifx\csname citenamefont\endcsname\relax
  \def\citenamefont#1{#1}\fi
\expandafter\ifx\csname url\endcsname\relax
  \def\url#1{\texttt{#1}}\fi
\expandafter\ifx\csname urlprefix\endcsname\relax\def\urlprefix{URL }\fi
\providecommand{\bibinfo}[2]{#2}
\providecommand{\eprint}[2][]{\url{#2}}

\bibitem[{\citenamefont{{Lentz} et~al.}({2011})\citenamefont{{Lentz},
  {Messer}, {Mezzacappa} et~al.}}]{opac}
\bibinfo{author}{\bibfnamefont{E.~J.} \bibnamefont{{Lentz}}},
  \bibinfo{author}{\bibfnamefont{O.~E.~B.} \bibnamefont{{Messer}}},
  \bibinfo{author}{\bibfnamefont{A.}~\bibnamefont{{Mezzacappa}}},
  \bibnamefont{et~al.}, \bibinfo{journal}{{in preparation}}
 (\bibinfo{year}{{2011}}).

\bibitem[{\citenamefont{{Lentz} et~al.}({2011})\citenamefont{{Lentz}, {Baird},
  {Hix} et~al.}}]{eos}
\bibinfo{author}{\bibfnamefont{E.~J.} \bibnamefont{{Lentz}}},
  \bibinfo{author}{\bibfnamefont{M.~L.} \bibnamefont{{Baird}}},
  \bibinfo{author}{\bibfnamefont{W.~R.} \bibnamefont{{Hix}}},
  \bibnamefont{et~al.}, \bibinfo{journal}{{in preparation}}
 (\bibinfo{year}{{2011}}).

\bibitem[{\citenamefont{Mezzacappa and Bruenn}(1993)}]{MeBr93a}
\bibinfo{author}{\bibfnamefont{A.}~\bibnamefont{Mezzacappa}} \bibnamefont{and}
  \bibinfo{author}{\bibfnamefont{S.~W.} \bibnamefont{Bruenn}},
  \bibinfo{journal}{ApJ} \textbf{\bibinfo{volume}{405}}, \bibinfo{pages}{637}
  (\bibinfo{year}{1993}).

\bibitem[{\citenamefont{{Liebend{\"o}rfer}
  et~al.}(2004)\citenamefont{{Liebend{\"o}rfer}, {Messer}, {Mezzacappa},
  {Bruenn}, {Cardall}, and {Thielemann}}}]{LiMeMe04}
\bibinfo{author}{\bibfnamefont{M.}~\bibnamefont{{Liebend{\"o}rfer}}},
  \bibinfo{author}{\bibfnamefont{O.~E.~B.} \bibnamefont{{Messer}}},
  \bibinfo{author}{\bibfnamefont{A.}~\bibnamefont{{Mezzacappa}}},
  \bibinfo{author}{\bibfnamefont{S.~W.} \bibnamefont{{Bruenn}}},
  \bibinfo{author}{\bibfnamefont{C.~Y.} \bibnamefont{{Cardall}}},
  \bibnamefont{and} \bibinfo{author}{\bibfnamefont{F.-K.}
  \bibnamefont{{Thielemann}}}, \bibinfo{journal}{ApJS}
  \textbf{\bibinfo{volume}{150}}, \bibinfo{pages}{263} (\bibinfo{year}{2004}).

\bibitem[{\citenamefont{Lattimer and Swesty}(1991)}]{LaSw91}
\bibinfo{author}{\bibfnamefont{J.}~\bibnamefont{Lattimer}} \bibnamefont{and}
  \bibinfo{author}{\bibfnamefont{F.~D.} \bibnamefont{Swesty}},
  \bibinfo{journal}{Nucl. Phys. A} \textbf{\bibinfo{volume}{535}},
  \bibinfo{pages}{331} (\bibinfo{year}{1991}).

\bibitem[{\citenamefont{{Woosley} and {Heger}}(2007)}]{WoHe07}
\bibinfo{author}{\bibfnamefont{S.~E.} \bibnamefont{{Woosley}}}
  \bibnamefont{and} \bibinfo{author}{\bibfnamefont{A.}~\bibnamefont{{Heger}}},
  \bibinfo{journal}{Phys. Rep.} \textbf{\bibinfo{volume}{442}},
  \bibinfo{pages}{269} (\bibinfo{year}{2007}).

\bibitem[{\citenamefont{Bruenn}(1985)}]{Brue85}
\bibinfo{author}{\bibfnamefont{S.~W.} \bibnamefont{Bruenn}},
  \bibinfo{journal}{ApJS} \textbf{\bibinfo{volume}{58}}, \bibinfo{pages}{771}
  (\bibinfo{year}{1985}).

\bibitem[{\citenamefont{{Hannestad} and {Raffelt}}(1998)}]{HaRa98}
\bibinfo{author}{\bibfnamefont{S.}~\bibnamefont{{Hannestad}}} \bibnamefont{and}
  \bibinfo{author}{\bibfnamefont{G.}~\bibnamefont{{Raffelt}}},
  \bibinfo{journal}{ApJ} \textbf{\bibinfo{volume}{507}}, \bibinfo{pages}{339}
  (\bibinfo{year}{1998}).

\bibitem[{\citenamefont{{Buras} et~al.}(2003)\citenamefont{{Buras}, {Janka},
  {Keil}, {Raffelt}, and {Rampp}}}]{BuJaKe03}
\bibinfo{author}{\bibfnamefont{R.}~\bibnamefont{{Buras}}},
  \bibinfo{author}{\bibfnamefont{H.}~\bibnamefont{{Janka}}},
  \bibinfo{author}{\bibfnamefont{M.~T.} \bibnamefont{{Keil}}},
  \bibinfo{author}{\bibfnamefont{G.~G.} \bibnamefont{{Raffelt}}},
  \bibnamefont{and} \bibinfo{author}{\bibfnamefont{M.}~\bibnamefont{{Rampp}}},
  \bibinfo{journal}{ApJ} \textbf{\bibinfo{volume}{587}}, \bibinfo{pages}{320}
  (\bibinfo{year}{2003}).

\bibitem[{\citenamefont{{Langanke} and {Mart{\'\i}nez-Pinedo}}(2000)}]{LaMa00}
\bibinfo{author}{\bibfnamefont{K.}~\bibnamefont{{Langanke}}} \bibnamefont{and}
  \bibinfo{author}{\bibfnamefont{G.}~\bibnamefont{{Mart{\'\i}nez-Pinedo}}},
  \bibinfo{journal}{Nucl. Phys. A} \textbf{\bibinfo{volume}{673}},
  \bibinfo{pages}{481} (\bibinfo{year}{2000}).

\bibitem[{\citenamefont{{Langanke} et~al.}(2003)\citenamefont{{Langanke},
  {Mart{\'\i}nez-Pinedo}, {Sampaio}, {Dean}, {Hix}, {Messer}, {Mezzacappa},
  {Liebend{\"o}rfer}, {Janka}, and {Rampp}}}]{LaMaSa03}
\bibinfo{author}{\bibfnamefont{K.}~\bibnamefont{{Langanke}}},
  \bibinfo{author}{\bibfnamefont{G.}~\bibnamefont{{Mart{\'\i}nez-Pinedo}}},
  \bibinfo{author}{\bibfnamefont{J.~M.} \bibnamefont{{Sampaio}}},
\bibnamefont{et~al.},
  \bibinfo{journal}{Phys. Rev. Lett.} \textbf{\bibinfo{volume}{90}},
  \bibinfo{pages}{241102} (\bibinfo{year}{2003}).

\bibitem[{\citenamefont{{Hix} et~al.}(2003)\citenamefont{{Hix}, {Messer},
  {Mezzacappa}, {Liebendoerfer}, {Sampaio}, {Langanke}, {Dean}, and
  {Martinez-Pinedo}}}]{HiMeMe03}
\bibinfo{author}{\bibfnamefont{W.~R.} \bibnamefont{{Hix}}},
  \bibinfo{author}{\bibfnamefont{O.~E.~B.} \bibnamefont{{Messer}}},
  \bibinfo{author}{\bibfnamefont{A.}~\bibnamefont{{Mezzacappa}}},
  \bibnamefont{et~al.},
  \bibinfo{journal}{Phys. Rev. Lett.} \textbf{\bibinfo{volume}{91}},
  \bibinfo{pages}{201102} (\bibinfo{year}{2003}).

\bibitem[{\citenamefont{{Burrows} and {Sawyer}}(1998)}]{BuSa98}
\bibinfo{author}{\bibfnamefont{A.}~\bibnamefont{{Burrows}}} \bibnamefont{and}
  \bibinfo{author}{\bibfnamefont{R.~F.} \bibnamefont{{Sawyer}}},
  \bibinfo{journal}{Phys. Rev. C} \textbf{\bibinfo{volume}{58}},
  \bibinfo{pages}{554} (\bibinfo{year}{1998}).

\bibitem[{\citenamefont{{Reddy} et~al.}(1998)\citenamefont{{Reddy}, {Prakash},
  and {Lattimer}}}]{RePrLa98}
\bibinfo{author}{\bibfnamefont{S.}~\bibnamefont{{Reddy}}},
  \bibinfo{author}{\bibfnamefont{M.}~\bibnamefont{{Prakash}}},
  \bibnamefont{and} \bibinfo{author}{\bibfnamefont{J.~M.}
  \bibnamefont{{Lattimer}}}, \bibinfo{journal}{Phys. Rev. D}
  \textbf{\bibinfo{volume}{58}}, \bibinfo{pages}{013009}
  (\bibinfo{year}{1998}).

\bibitem[{\citenamefont{{Horowitz}}(2002)}]{Horo02}
\bibinfo{author}{\bibfnamefont{C.~J.} \bibnamefont{{Horowitz}}},
  \bibinfo{journal}{Phys. Rev. D} \textbf{\bibinfo{volume}{65}},
  \bibinfo{pages}{43001} (\bibinfo{year}{2002}).

\bibitem[{\citenamefont{{Rampp} et~al.}(2002)\citenamefont{{Rampp}, {Buras},
  {Janka}, and {Raffelt}}}]{RaBuJa02}
\bibinfo{author}{\bibfnamefont{M.}~\bibnamefont{{Rampp}}},
  \bibinfo{author}{\bibfnamefont{R.}~\bibnamefont{{Buras}}},
  \bibinfo{author}{\bibfnamefont{H.}~\bibnamefont{{Janka}}}, \bibnamefont{and}
  \bibinfo{author}{\bibfnamefont{G.}~\bibnamefont{{Raffelt}}}, in
  \emph{\bibinfo{booktitle}{Nuclear Astrophysics}}, edited by
  \bibinfo{editor}{\bibnamefont{{W.~Hillebrandt \& E.~M{\"u}ller}}}
  (\bibinfo{year}{2002}), pp. \bibinfo{pages}{119--125}.

\bibitem[{\citenamefont{{Shen} et~al.}(1998)\citenamefont{{Shen}, {Toki},
  {Oyamatsu}, and {Sumiyoshi}}}]{ShToOy98}
\bibinfo{author}{\bibfnamefont{H.}~\bibnamefont{{Shen}}},
  \bibinfo{author}{\bibfnamefont{H.}~\bibnamefont{{Toki}}},
  \bibinfo{author}{\bibfnamefont{K.}~\bibnamefont{{Oyamatsu}}},
  \bibnamefont{and}
  \bibinfo{author}{\bibfnamefont{K.}~\bibnamefont{{Sumiyoshi}}},
  \bibinfo{journal}{Prog. Theor. Phys.} \textbf{\bibinfo{volume}{100}},
  \bibinfo{pages}{1013} (\bibinfo{year}{1998}).

\bibitem[{\citenamefont{{Swesty} et~al.}(1994)\citenamefont{{Swesty},
  {Lattimer}, and {Myra}}}]{SwLaMy94}
\bibinfo{author}{\bibfnamefont{F.~D.} \bibnamefont{{Swesty}}},
  \bibinfo{author}{\bibfnamefont{J.~M.} \bibnamefont{{Lattimer}}},
  \bibnamefont{and} \bibinfo{author}{\bibfnamefont{E.~S.}
  \bibnamefont{{Myra}}}, \bibinfo{journal}{ApJ} \textbf{\bibinfo{volume}{425}},
  \bibinfo{pages}{195} (\bibinfo{year}{1994}).

\bibitem[{\citenamefont{{Hix} et~al.}(2008)\citenamefont{{Hix}, {Lentz},
  {Baird}, {Messer}, and {Mezzacappa}}}]{HiLeBa08}
\bibinfo{author}{\bibfnamefont{W.~R.} \bibnamefont{{Hix}}},
  \bibinfo{author}{\bibfnamefont{E.~J.} \bibnamefont{{Lentz}}},
  \bibinfo{author}{\bibfnamefont{M.~L.} \bibnamefont{{Baird}}},
  \bibinfo{author}{\bibfnamefont{O.~E.~B.} \bibnamefont{{Messer}}},
  \bibnamefont{and}
  \bibinfo{author}{\bibfnamefont{A.}~\bibnamefont{{Mezzacappa}}}, in
  \emph{\bibinfo{booktitle}{Nuclei in the Cosmos (NIC X)}}, 
  \bibinfo{number}{{PoS(NIC X)017}}
  (\bibinfo{year}{2008}).

\bibitem[{\citenamefont{{Janka} et~al.}(2005)\citenamefont{{Janka}, {Buras},
  {Kitaura Joyanes}, {Marek}, {Rampp}, and {Scheck}}}]{JaBuKi05}
\bibinfo{author}{\bibfnamefont{H.-T.} \bibnamefont{{Janka}}},
  \bibinfo{author}{\bibfnamefont{R.}~\bibnamefont{{Buras}}},
  \bibinfo{author}{\bibfnamefont{F.~S.} \bibnamefont{{Kitaura Joyanes}}},
  \bibinfo{author}{\bibfnamefont{A.}~\bibnamefont{{Marek}}},
  \bibinfo{author}{\bibfnamefont{M.}~\bibnamefont{{Rampp}}}, \bibnamefont{and}
  \bibinfo{author}{\bibfnamefont{L.}~\bibnamefont{{Scheck}}},
  \bibinfo{journal}{Nucl. Phys. A} \textbf{\bibinfo{volume}{758}},
  \bibinfo{pages}{19} (\bibinfo{year}{2005}).

\bibitem[{\citenamefont{{Marek}}(2007)}]{Mare07}
\bibinfo{author}{\bibfnamefont{A.}~\bibnamefont{{Marek}}}, Ph.D. thesis,
  \bibinfo{school}{{Technische Universit{\"a}t M{\"u}nchen}}
  (\bibinfo{year}{2007}).

\bibitem[{\citenamefont{{Sumiyoshi} et~al.}(2005)\citenamefont{{Sumiyoshi},
  {Yamada}, {Suzuki}, {Shen}, {Chiba}, and {Toki}}}]{SuYaSu05}
\bibinfo{author}{\bibfnamefont{K.}~\bibnamefont{{Sumiyoshi}}},
  \bibinfo{author}{\bibfnamefont{S.}~\bibnamefont{{Yamada}}},
  \bibinfo{author}{\bibfnamefont{H.}~\bibnamefont{{Suzuki}}},
  \bibinfo{author}{\bibfnamefont{H.}~\bibnamefont{{Shen}}},
  \bibinfo{author}{\bibfnamefont{S.}~\bibnamefont{{Chiba}}}, \bibnamefont{and}
  \bibinfo{author}{\bibfnamefont{H.}~\bibnamefont{{Toki}}},
  \bibinfo{journal}{ApJ} \textbf{\bibinfo{volume}{629}}, \bibinfo{pages}{922}
  (\bibinfo{year}{2005}).

\end{thebibliography}

%\end{thebibliography}

\end{document}